%% file: main.tex
\documentclass[conference]{IEEEtran}
\IEEEoverridecommandlockouts
\usepackage{cite}
\usepackage{amsmath,amssymb,amsfonts}
\usepackage{algorithmic}
\usepackage{graphicx}
\usepackage{textcomp}
\usepackage[dvipsnames]{xcolor}

\usepackage{pgf-pie}                                
\usepackage{siunitx}
\usepackage{hyperref}
\usepackage[nolist]{acronym}
\usepackage[flushleft]{threeparttable}
\usepackage{placeins}
\usepackage{svg}                                    

\def\BibTeX{{\rm B\kern-.05em{\sc i\kern-.025em b}\kern-.08em
    T\kern-.1667em\lower.7ex\hbox{E}\kern-.125emX}}

\usepackage{eso-pic}
\usepackage{url}
\AddToShipoutPictureBG*{
  \AtPageUpperLeft{%
    \put(0,-40){\raisebox{15pt}{\makebox[\paperwidth]{\begin{minipage}{21cm}\centering
      \textcolor{gray}{This article has been accepted for publication in the proceedings of the 2023 9th IEEE International Workshop on Advances in Sensors \\and Interfaces (IWASI). Content may change prior to final publication. DOI: 10.1109/IWASI58316.2023.10164426} 
    \end{minipage}}}}%
  }
  \AtPageLowerLeft{%
    \raisebox{25pt}{\makebox[\paperwidth]{\begin{minipage}{21cm}\centering
      \textcolor{gray}{ \copyright 2023 IEEE.  Personal use of this material is permitted.  Permission from IEEE must be obtained for all other uses, in any current or future media, including reprinting/republishing this material for advertising or promotional purposes, creating new collective works, for resale or redistribution to servers or lists, or reuse of any copyrighted component of this work in other works.
      }
    \end{minipage}}}%
  }
}

\begin{document}
\newcommand{\todo}[1]{\textcolor{red}{#1}}                          
\newcommand{\done}[1]{\textcolor{Green}{#1}}                        
\DeclareSIUnit{\dBm}{\text{dBm}}                                    
\newcommand*{\rom}[1]{\expandafter\@slowromancap\romannumeral #1@}  

\title{A LoRa-based Energy-efficient Sensing System for Urban Data Collection}

\author{
\IEEEauthorblockN{
Lukas Schulthess\IEEEauthorrefmark{1}, 
Tiago Salzmann\IEEEauthorrefmark{1},
Christian Vogt\IEEEauthorrefmark{1},
Michele Magno\IEEEauthorrefmark{1}}
\IEEEauthorblockA{\IEEEauthorrefmark{1}Dept. of Information Technology and Electrical Engineering, ETH Z\"{u}rich, Switzerland} 
}

\maketitle

\input{src/00_Abstract}

\vspace{10pt}

\input{src/01_Introduction}
\input{src/03_Architecture}
\input{src/04_Results}
\input{src/05_Conclusion}
\input{inc/acronyms}
\bibliographystyle{IEEEtran} 
\bibliography{main}

\end{document}

%% file: src/00_Abstract.tex
\begin{abstract}
Nowadays, cities provide much more than shopping opportunities or working spaces. Individual locations such as parks and squares are used as meeting points and local recreation areas by many people. 
To ensure that they remain attractive in the future, the design of such squares must be regularly adapted to the needs of the public. These utilization trends can be derived using public data collection. The more diverse and rich the data sets are, the easier it is to optimize public space design through data analysis. Traditional data collection methods such as questionnaires, observations, or videos are either labor intensive or cannot guarantee to preserve the individual's privacy.

This work presents a privacy-preserving, low-power, and low-cost smart sensing system that is capable of anonymously collecting data about public space utilization by analyzing the occupancy distribution of public seating. 
To support future urban planning the sensor nodes are capable of monitoring environmental noise, chair utilization, and their position, temperature, and humidity and provide them over a city-wide \ac{LoRaWAN}. The final sensing system's robust operation is proven in a trial run at two public squares in a city with 16 sensor nodes over a duration of two months. By consuming 33.65 mWh per day with all subsystems enabled, including sitting detection based on a continuous acceleration measurement operating on a robust and simple threshold algorithm, the custom-designed sensor node achieves continuous monitoring during the 2-month trial run. The evaluation of the experimental results clearly shows how the two locations are used, which confirms the practicability of the proposed solution. 
All data collected during the field trial is publicly available as open data.

\end{abstract}

%% file: src/01_Introduction.tex
\section{Introduction}\label{sec:introduction}

The concept of smart cities influences the operation and organization of several service domains of today's urban areas \cite{yang_smart_city_development_2021}. 
With the help of urban computing - the concept of applying technological solutions in a public environment - resource management, transportation, and infrastructure maintenance can be optimized and simplified \cite{mulero_ressource_planning_2020, kong_bus_profiling_2020}.
But not only already existing infrastructure is affected by this paradigm; city planners use data on how much the public city spaces are used and how their attractiveness can be improved, for example by providing extra sitting opportunities and artificial shading \cite{mazhar2020characterizing}. 
Such data contain important information, which serves as a basis for future public space design. 
This process of improving the city's public service and future room planning by acquiring, integrating, and analyzing data is called urban planning \cite{li2018planning}. 
It relies on combined data from multiple sources; movement by pedestrians, cyclists, and public transport counts, but also utilizes other information sources, such as environmental measurements \cite{kaginalkar_review_2021}, noise \cite{peckens_wireless_2018} or energy consumption \cite{thorve_privacy_2018}. 
Many locations are suitable for gathering those data, however, the location and the type of data may depend on the topic of interest.
Focusing on the improvement of public spaces, simple approaches like pedestrian counting \cite{de_ruyter_automatic_2010} can already indicate how often a certain location is visited. Many experimental as well as commercial sensing systems already allow for measuring pedestrian utilization at fixed positions in a city \cite{akhter_iot_2019, akhter_design_2019}.
In addition to people counting, data originating from social media or cell phones can be utilized as a means to assess the usage of public spaces \cite{ciuccarelli_visualizing_2014}. It has the benefit of encompassing the entire metropolis rather than being restricted to a single area.
While the latter two approaches provide reliable insights into the usage of city spaces such as shops, stairways, and elevators over time, the usage of public infrastructure like benches or chairs, but also shaded places cannot be precisely measured. \\
Installing local sensor nodes for specific sensing tasks could complement already existing solutions in specific areas \cite{lau_sensor_fusion_monitoring_2018}. 
For example, by leveraging wireless communication \cite{rashid_applications_2016, carminati_prospects_2019}, only people being present at squares are accounted for,  but not if they use the provided seating, prefer sunny places or shaded ones or prefer loud or quiet environments. 
Questionnaires and local observations can gather those missing measurement values, but they are time-consuming and labor-intensive, and do not provide constant data over a long period of time. 

The \ac{IoT} is changing the world, aiming to bring invisible and innovative solutions to connect digital sensors devices with the internet \cite{mehmood2017internet}. 
Small, low-power and intelligent smart sensors are today's reality thanks to the technological advancement in major electronic fields such as wireless communication, sensing, processing, and power-efficient \acp{MCU}.
On the other hand, building and deploying billions of devices covering a wide range of applications is bringing many research challenges such as energy management on small batteries \cite{said_iot_energy_management_2020}, security and privacy \cite{li_iot_privacy_2019}, energy efficient processing and communication \cite{yosuf_iot_processing_2020}, localization and many others. Tackling these challenges drives research into understanding how smart and unobtrusive \ac{IoT} devices require to be designed, adopted, and deployed in real field \cite{zhang_iot_unobstrusive_2016}. 

Well-positioned smart and inconspicuous \ac{IoT} sensor nodes can facilitate the process of data acquisition and improve the data quality and richness of current approaches by continuously observing public spaces and offering quantitative data on usage of infrastructure, for example, public chairs \cite{he_intelligent_chair_system_2016, krejcar_smart_furniture_2019}. In contrast to video surveillance, such localized sensor nodes can preserve privacy and thus also improve acceptance in public \cite{eckhoff_privacy_smart_cities_2018}.
In addition to the privacy-preserving aspect, these sensor nodes should be designed with low cost in mind for both device procurement and maintenance in order to enable future large-scale deployments. As an example, the primary component of maintenance work for such sensor nodes is the replacement or recharging of batteries \cite{ma_eh_iot_2020}. In the case of operating expenses, it is crucial to consider energy efficiency when designing \ac{IoT} devices, including optimizations in both software and hardware. For instance, a carefully selected communication technology further reduces the sensor nodes' power consumption during data transmission. Finally, \ac{IoT} devices for smart cities must be durable and able to resist the environmental factors that will affect their deployment. While studies frequently take place in controlled settings, this may not accurately represent how they behave in actual installations in urban areas. Therefore, conducting an on-site assessment is essential.

This paper presents the design and implementation of a LoRa-based sensing system for chair monitoring and urban data collection in public spaces.
It features built-in environmental data collection and processing capabilities, onboard classification of the chair's occupancy, and a back-end for additional statistical data recording. 
This sensor node is built into waterproof and robust housing to withstand weather conditions in urban areas. 
The proposed system has been successfully tested by collecting environmental data such as temperature, humidity, ambient noise, GNSS position, and general occupancy of the chairs over a time period of two months at two different public squares in the city.
The sensor node can reach a battery lifetime of 2 months if all the subsystems are active, which reduces maintenance requirements and enables quick implementation.
By deactivating the sensor nodes' GNSS, a battery lifetime of up to 5 months can be achieved.


%% file: src/03_Architecture.tex

\section{System Definition}\label{sec:system_architecture}
\begin{figure}[]
    \includegraphics[width=\columnwidth]{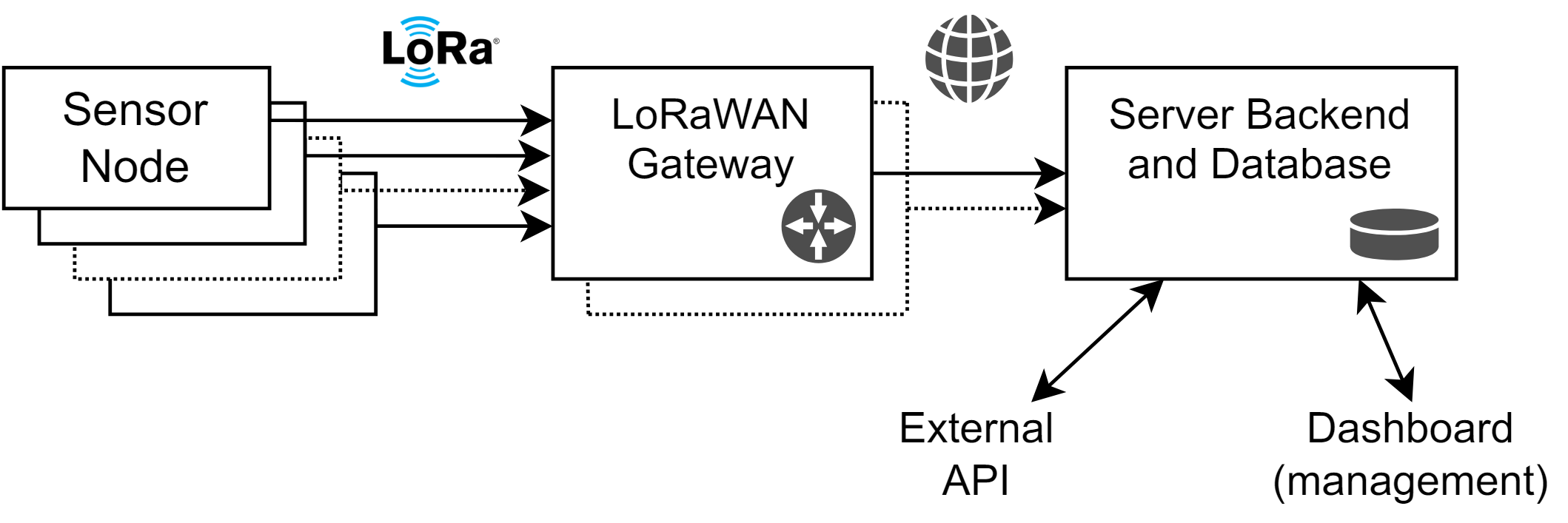}
    \caption{System-wide overview of multiple sensor nodes connected to the server back-end over \ac{LoRaWAN} gateways.}
    \label{fig:system_wide_architecture}
\end{figure}

The proposed sensing system for automated data acquisition consists of both, hardware and software components.
Deployed sensor nodes act as edge devices that collect, preprocess, and forward the measurements using \ac{LoRaWAN} to dedicated gateways. The gateways are operated by a commercial vendor and were already available and used for other smart sensors located in the city. From there, the received packets are forwarded to a central server, which decodes the payload and stores the measurements in a MySQL database. Furthermore, it offers a simple management dashboard and graphical measurement overview, as well as external API access to the measurements in the database, \autoref{fig:system_wide_architecture}. \newline
The collected data and measurement system was evaluated within the city administration to conform with local privacy and data protection regulations.

\subsection{Hardware Architecture}\label{sec:hardware_architecture}
The sensor node has been designed to collect specific environmental data that can be used to draw conclusions about the squares utilization and to facilitate the measurement setup:
\begin{itemize}
    \item \textit{Accelerometer:} It allows recognizing physical events, such as chair movement, sitting down, and standing up. From this, generalized statements like the chair's occupancy can be derived. 
    \item \textit{Microphone:} Measuring the noise level can give more insights how the specific location is used. Loud sounds could indicate a festival or construction work.
    \item \textit{Temperature- and humidity sensor:} Tracking the temperature and humidity on every chair provides a fine-grained dataset from which information about the individual chair's position can be derived. Locations in the shade tend to have lower temperatures than sunny places.
    \item \textit{GNSS module:} Having the GNSS position of each chair facilitates the experimental setup. Chairs can be automatically assigned to a square at the server backend. Further, lost or stolen devices can be tracked and recovered again.
\end{itemize}

\begin{figure}[!b]
    \includegraphics[width=\columnwidth]{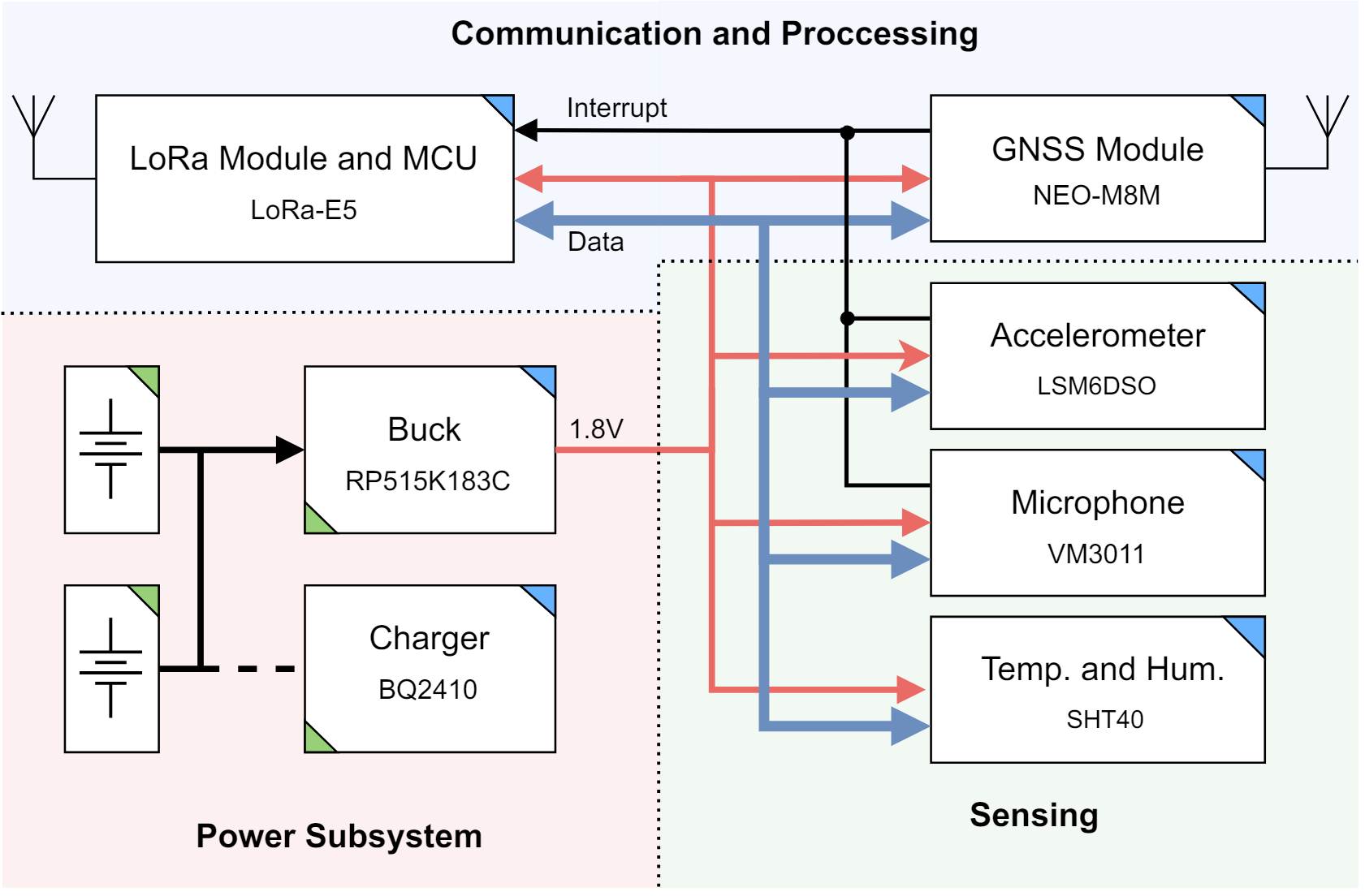}
    \caption{High-level architecture of the proposed sensor node, divided into the three main sections communication and processing (blue), 
    environmental sensing (green, and power distribution (red).}
    \label{fig:high_level_hardware}
\end{figure}

The designed sensor node can be divided into three main parts: communication and processing, environmental sensing, and power distribution, \autoref{fig:high_level_hardware}. Commercial off-the-shelf components reduce the overall cost per node while giving flexibility for custom adaptations.

\textit{Communication and Processing:} 
The sensor node is controlled by an ultra-low power multi-modulation wireless STM32\-WLE5JC microcontroller from ST Microelectronics.
Based on an ARM Cortex-M4 and the sub-GHz radio SX126x from Semtech, the \ac{SoC} is responsible to collect data from the sensors, doing onboard processing, preparing and transmitting LoRa packets. 
By choosing a Seeed Studio LoRa-E5 module, which already integrates the RF-matching, the overall system design is further simplified.
LoRa packets are transmitted over an Antenova SR42I010-R PCB antenna.
\ac{GNSS} data is acquired by an uBlox NEO-M8M GNSS module using the passive PCB antenna SR4G008 from Antenova.

\textit{Environmental Sensing:} 
The following sensors have been implemented to measure environmental conditions:
For measuring the chair's occupancy, the inertial measurement unit \ac{IMU} LSM6DSO from ST Microelectronics has been selected. Further, to determine the sitting comfort based on environmental influences, a Sensirion STH40 humidity and temperature sensor for sensing the weather conditions and a Vesper VM3011 MEMS microphone to determine the average noise exposure have been integrated. The required opening in the waterproof housing is sealed by a Sensirion SF2 filter cap that protects the sensor and the internal circuitry against dust particles and water immersion.

\textit{Power and Subsystem:} 
The device can be powered either via two rechargeable or non-rechargeable batteries. 
By selecting the appropriate configuration, the batteries are set in series or in parallel to each other. 
For non-rechargeable standard AA alkaline batteries, the series configuration is needed to provide a higher battery voltage than the systems voltage of \SI{1.8}{\volt}. 
By configuring the batteries in parallel, the non-rechargeable batteries are replaced with rechargeable ones. 
The single-cell li-ion battery charger BQ24210DQCR from Texas Instruments allows charging from external power sources using the onboard USB-C connector.
In both cases, the system voltage of \SI{1.8}{\volt} is generated from the battery voltage using a Nisshinbo RP515K183C buck converter with integrated battery level measurement.  

\subsection{Firmware}
\label{sect:sysfirmware}
\begin{figure}[]
    \includegraphics[width=\columnwidth]{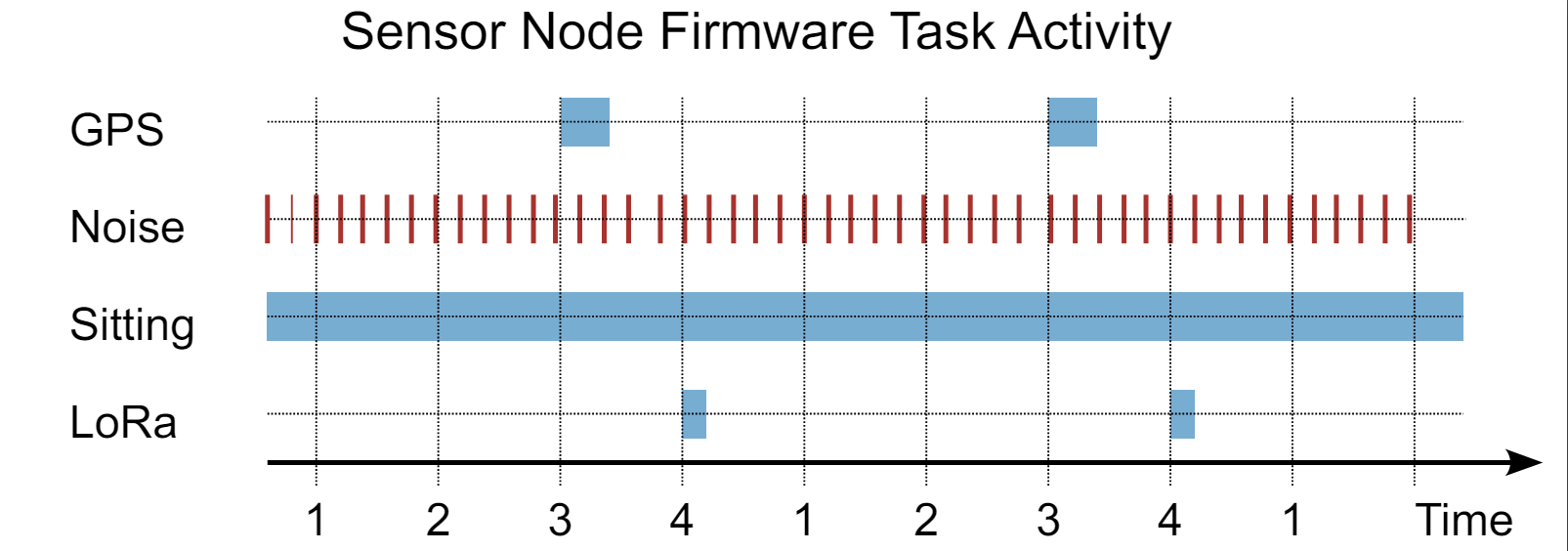}
    \caption{Task activity of the sensor node, arranged according to sampling intervals (1-4). At sampling interval 4, temperature and humidity are sampled and transmitted together with all other measurements via \ac{LoRaWAN}.}
    \label{fig:sequence_tasks}
\end{figure}

The firmware is based on Zephyr \ac{RTOS} Version 3.1.99\footnote{\url{https://www.zephyrproject.org/}}
and is organized into 4 threads: (I) a main thread handling \ac{LoRaWAN} communication and the fast sensor readings (e.g. temperature/humidity), (II) a noise detection thread that handles the sampling of the noise level from the microphone in \SI{1}{\second} intervals and averaging it, (III) a sitting detection thread, sampling the accelerometer and deciding if a person is sitting on the chair and (IV) a \ac{GNSS} thread to control the \ac{GNSS} module. \newline
The power consuming task of localization over \ac{GNSS} is repeated every 2 hours. For occupancy, a more fine-grained data collection was desired, thus the sequence of these tasks is split into 4 repeating sampling intervals, each \SI{30}{\minute} long, as shown in \autoref{fig:sequence_tasks}. 
\ac{GNSS} is active during the second-last sampling interval before the \ac{LoRaWAN} transmission, as the time needed for the position fix to occur is not known prior to the measurement. 
The noise level and sitting duration are averaged over each sampling interval and are transmitted over \ac{LoRa} to reduce the amount of transmitted data. Temperature and humidity are sampled once immediately before the \ac{LoRa} transmission and packed into the data frame, together with the rest of the measurement values. 
The payload of the data packet is composed of 29 bytes in total.
It consists of a 1-byte header to specify the payload structure, 1 byte for debug information, 14 bytes for \ac{GNSS} position (4 bytes each for latitude, longitude and time, and 2 bytes for the position accuracy), 1-byte battery level, 2-byte temperature, 2-byte humidity, and four sitting and noise values (each 1 byte)  for the sampling intervals. For this 29 bytes payload, the air-time of the packet is estimated to be \SI{1.24}{\second} for SF12 and \SI{75}{\milli\second} for SF7.
The data transmission over \ac{LoRaWAN} is with $\approx$ \SI{110}{\milli\watt} the most power-hungry part of the designed hardware, and its air-time should be as short as possible. 
The air-time calculation for \ac{LoRaWAN} transmissions is defined by the settings of the \ac{LoRaWAN} (such as spreading factor, bandwidth etc.) and the payload size \cite{semtech_corporation_an120013_2013}. Thus, data transmissions are set after 4 intervals to reduce the additional overhead by the LoRaWAN header.

To save energy on the sensor node, the power-intensive tasks (\ac{GNSS} and \ac{LoRa}) are duty-cycled, with the firmware managing the activation of the sensors by power-gating. 
An overview of the tasks and their activation and execution duration is given in \autoref{tab:task_times}. 
The total power consumption of \ac{GNSS} localization and \ac{LoRa} data packet transmission depends on the interval and was set to occur every 2 hours as a trade-off between power efficiency and real-time sensing.\newline 
The maximum duration of the \ac{GNSS} modules activity is limited to \SI{5}{\minute} to reduce the worst-case power draw. 
In contrast, the noise and sitting detection is significantly faster, as they require continuous recording of the sensors to also capture sparse events. 
Sitting detection is continuously operating on data sampled by the accelerometer.
Environmental noise in dBSPL is directly detected by the microphone itself providing a \SI{1}{\second} average that is read over I2C every 1 second.

\begin{table}[]
    \centering
    \caption{Overview of task execution time as well as activation periods in realized sensor node firmware}
    \label{tab:task_times}
    \begin{tabular}{lcc}\hline
       \textbf{Task} &  \textbf{Active Time [s/Call]} &  \textbf{Call Intervall [s]} \\\hline
       (I) Lora & 10.2 & 7200\\\hline
       (II) Noise & 0.88e-3 & 1 \\\hline
       (III) Sitting & 1.1e-3 & 1/26 \\\hline
       (IV) GNSS & max.300 &7200\\\hline
    \end{tabular}
\end{table}


%% file: src/04_Results.tex
\section{Experimental Results}\label{sec:results}
\begin{figure}[]
    \includegraphics[width=\columnwidth]{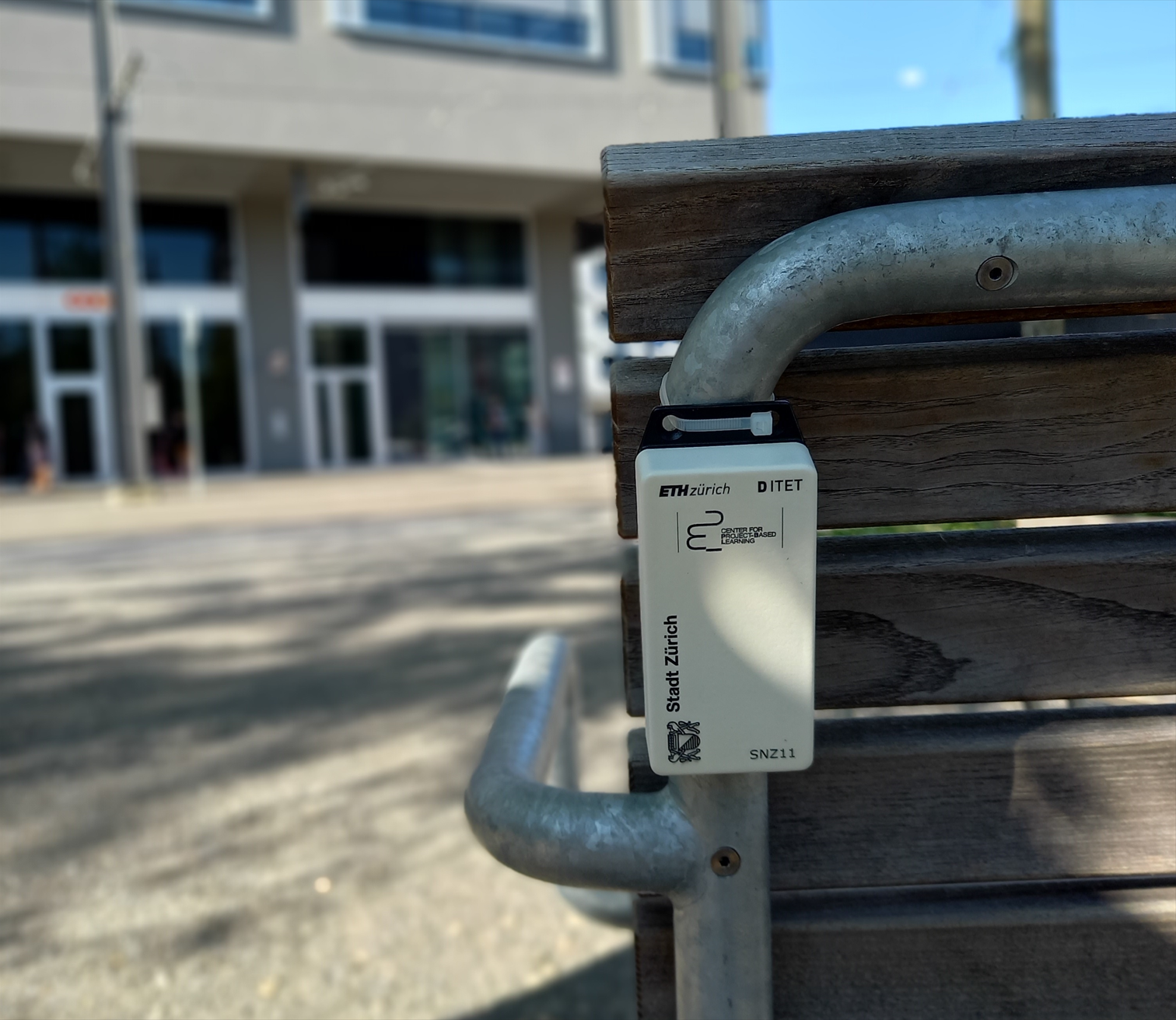}
    \caption{An installed sensor node attached to a public chair on an urban square. Experimental results have shown that cases with white coating have a lower internal temperature of approximately \SI{10}{\celsius} when compared to black casings. }
    \label{fig:sensor_depolyment}
\end{figure}

The proposed system has been evaluated in a real application scenario for 2 months. 
This section presents the final results that have been collected during this time on two different squares in a city. Square \emph{M} is located in the old part of the city center surrounded by stores and restaurants and has no greening. In the contrary, Square \emph{V} is located next to a train station in the suburban areas and has greened areas and trees. \newline
During the field test, 5 of the total 16 sensor nodes stopped working due to theft or vandalism.
The rest of the nodes worked without issue for the 2-month period of the test run. 
\autoref{fig:sensor_depolyment} shows how a sensor node has been attached to a chair for the experimental verification period using strong plastic zip ties. This connection method has been used deliberately, as it represents a good trade-off between a strong connection and still being simple to open. This is important in the context of vandalism to prevent severe damage to the chair.

\subsection{Power Consumption} \label{subsec:power_consumption}
To reduce overall power consumption, the design relies on power gating for the \ac{GNSS} module as well as for the temperature and humidity sensor. 
Furthermore, the microphone and \ac{IMU} were selected to have low standby and operating currents ($\approx$ \SI{10}{\micro\ampere} for the microphone in zero-power listening mode, and $\approx$ \SI{14}{\micro\ampere} for the \ac{IMU}).
A table of the individual tasks' power consumption is given in \autoref{tab:energy}. 
The node consumes approx. \SI{33.65}{\milli\watt\hour} per day when all subsystems are activated, allowing for an operation of 2 months on an alkaline battery with a capacity of approx. \SI{2000}{\milli\ampere\hour}. 

\begin{table}[]
    \centering
    \caption{Energy consumption of sensor node per day per task}
    \label{tab:energy}
    \begin{tabular}{ll}\hline
        {\bf Task} & {\bf Energy}\\ \hline
        Noise and Sitting &  \SI{9.24}{\milli\watt\hour} per day\\\hline
        GPS & \SI{22.5}{\milli\watt\hour} per day\\\hline
        LoRaWAN & \SI{1.91}{\milli\watt\hour} per day\\ \hline
        \textbf{Total} &\textbf{\SI{33.65}{\milli\watt\hour} per day} \\\hline
    \end{tabular}
\end{table}

\begin{figure}[]
    \includegraphics[width=\columnwidth]{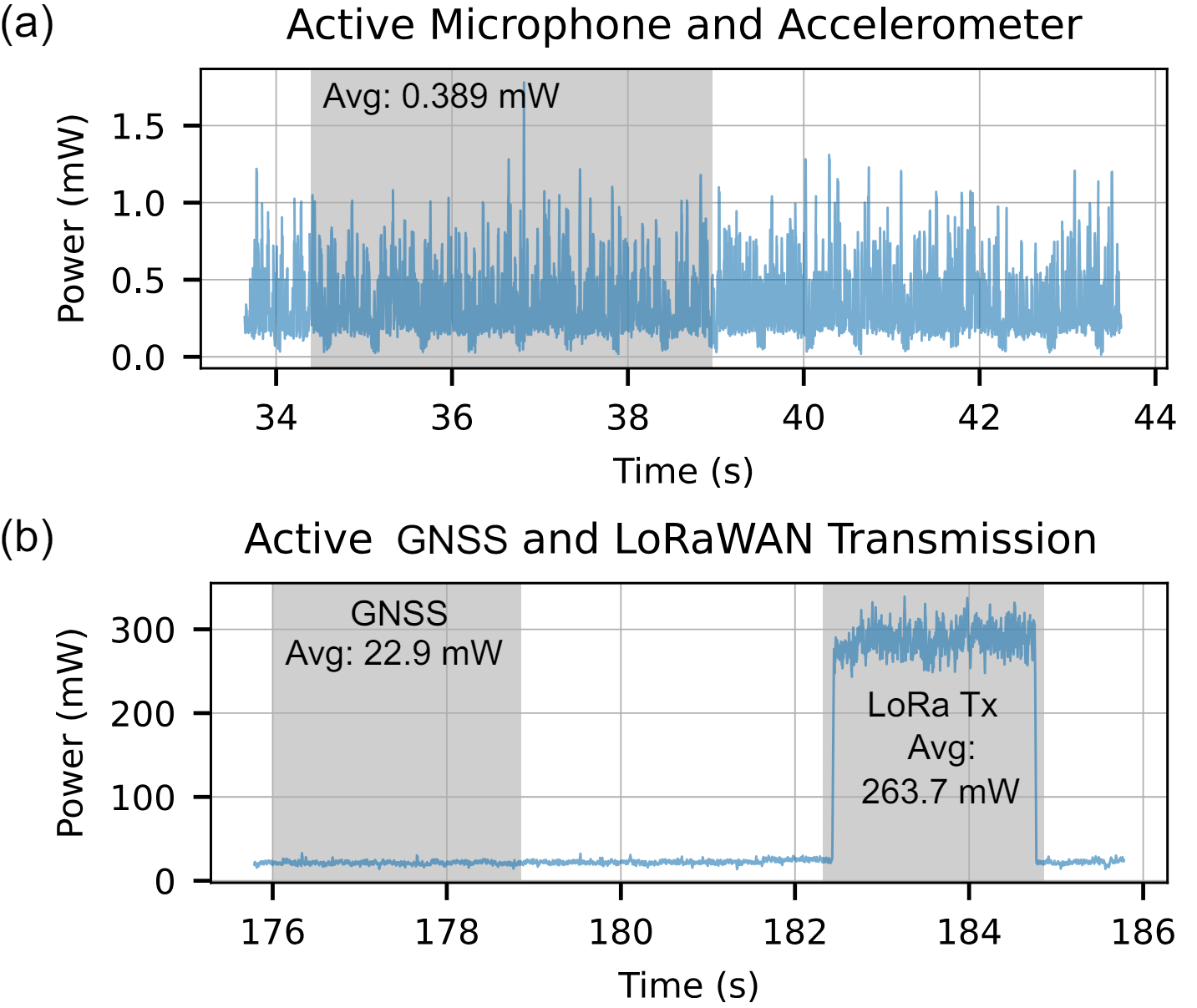}
    \caption{Power consumption of the designed sensor node over time in 3 different modes: Standby and sitting/noise detection (a) and GNSS recording and LoRaWAN transmission (b). For each operation mode, the average power draw (Avg) is given.}
    \label{fig:plot_power_over_time}
\end{figure}

In the worst case, with no or very poor GPS reception, the GPS module consumes with approximately 67\% of the total device energy a significant portion of the overall power (see \autoref{tab:energy}). 
The second largest energy consumer is the noise and sitting task with \SI{9.24}{\milli\watt\hour} per day, as the main \ac{MCU} is required to wake up every 1 second and process the accelerometers data as well as get the averaged microphone noise level. 
The lowest consumer is the \ac{LoRaWAN} transmission with SF12 and 29 bytes of payload at \SI{1.91}{\milli\watt\hour} per day. \newline
\autoref{fig:plot_power_over_time} presents an example of the power draw of the different tasks over time. 
It has been measured using a Nordic Semiconductor Power Profiler Kit II connected to the battery input of the sensor node at \SI{3}{\volt}. The continuous background task of sitting detection and noise recording consumes an average of $\approx$
\SI{0.39}{\milli\watt}, the duty-cycled tasks of \ac{GNSS} and \ac{LoRaWAN} consume $\approx$ \SI{22.9}{\milli\watt} and  $\approx$ \SI{263.7}{\milli\watt} respectively. The \ac{LoRaWAN} measurement is based on a spreading factor of 12 and the maximum permitted transmit power.

\subsection{Data combination and Evaluation}
\begin{figure}[]
\centering
    \includegraphics[width=0.96\columnwidth]{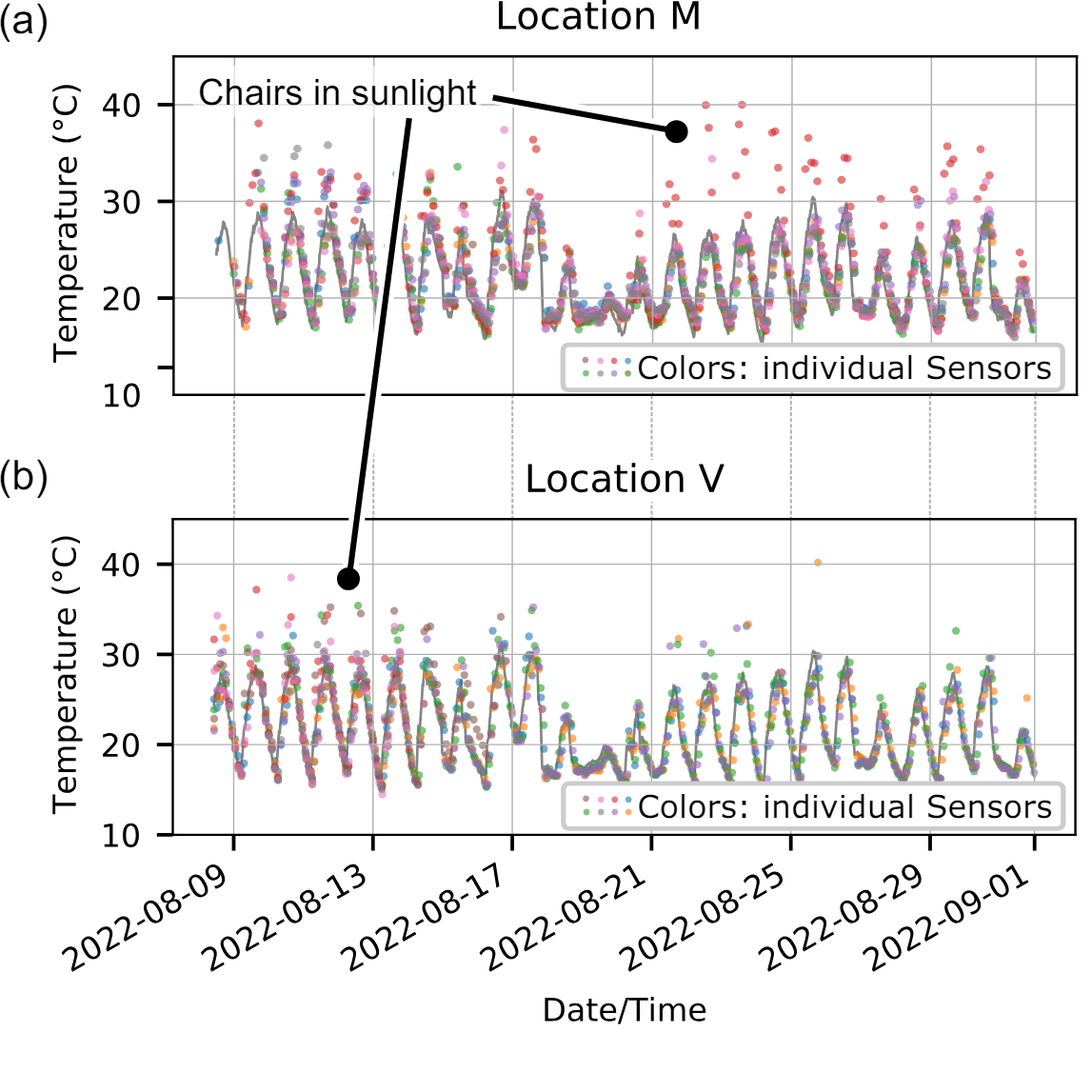}
    \caption{Temperature of the city's reference sensor (blue line) and the collected temperatures of the individual sensor (SNZxx) over time for the two squares (a,b) showing the ability to detect chairs in direct sunlight based on increased temperature with respect to reference sensor.}
    \label{fig:sunny}
\end{figure}

Utilizing existing public data from reference sensors in the same district as the sensors have been placed, the data collected by the sensor nodes can be augmented to give a better insight into public space utilization. 
A simple example is given over 2 weeks in \autoref{fig:sunny} on both test locations M and V. 
Day and night temperature cycles are clearly visible, ranging between $\approx$ \SI{15}{\celsius} and \SI{30}{\celsius}. 
During the night, all sensor nodes follow the reference sensor of the city. However, during the day some of the sensor measurements differ significantly with measured temperatures over \SI{40}{\celsius}. 
This is because sensor nodes placed in direct sunlight show a significantly higher temperature due to the housing's heating.
This allows measuring the environment of the sensor when compared with other temperature sensors of the city - for example, if the sensor node is placed in direct sunlight or shade.

Similarly, comparing humidity and seat occupation in \autoref{fig:hum_seat} yields the expected result of low seat occupation during measurement points with high humidity. Comparing the humidity measurement with rainfall, the sensor nodes record a humidity between 80\% RH and 100\% RH during rain. Therefore, high humidity values indicate rain or bad weather. 
Combining the humidity with the recorded seat occupation, the chair's usage reduces significantly during rainy weather, as indicated by no measurements being located in the upper right corners of \autoref{fig:hum_seat} (a) and \autoref{fig:hum_seat} (b). 
Comparing the two squares with each other, the city square of \autoref{fig:hum_seat} (a) shows a generally higher occupation irrespective of humidity than the square on the outskirts of the city \autoref{fig:hum_seat} (b).

\begin{figure}[]
    \includegraphics[width=0.97\columnwidth]{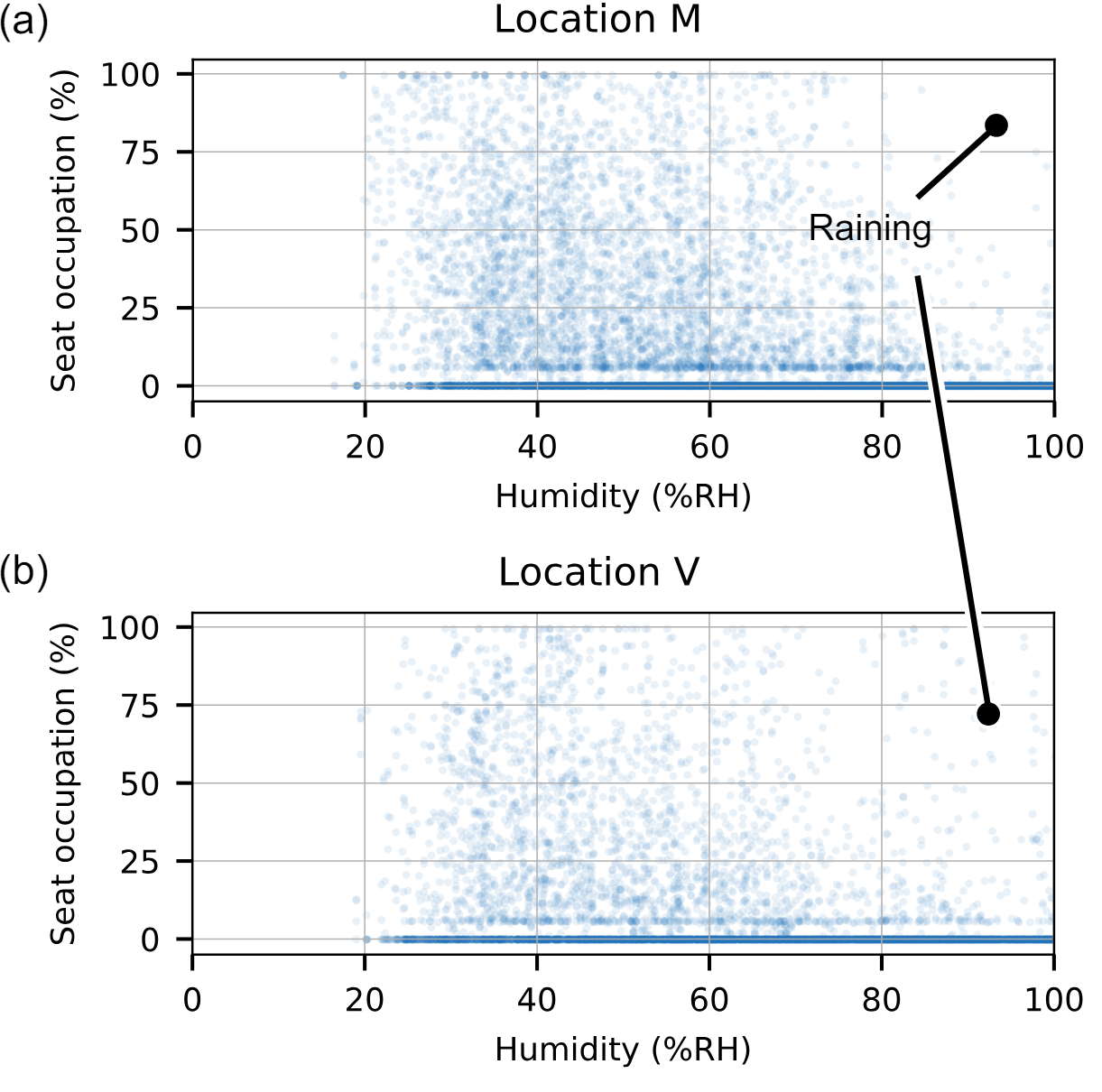}
    \caption{Recorded humidity (an indicator of rain) versus the seat occupation on the two locations in the city (a,b) with each dot representing a measurement value of a sensor over the experiment duration. High humidity values correlate with less occupation.}
    \label{fig:hum_seat}
\end{figure}

When comparing the accumulated sitting time per day and square with the city's official temperature recordings, a clear tendency can be derived.
During warm days (above \SI{20}{\celsius}), the centrally located square \emph{M} is more used than square \emph{V}. during cold days (below \SI{15}{\celsius}), the chairs are no longer used on \emph{M} whereas for \emph{V} the average sitting time decreases, \autoref{fig:hum_seat} (a).
The reason for this lies in the different locations and environments: \emph{M} is located in the city center and attractive to visitors and free-time activities in cafes and restaurants, especially on warm and sunny days.
Square \emph{V} is located in the suburban area and this is mainly attractive for commuters. This can be seen from the accumulated sitting time; In total, it is less actively used compared to \emph{M} but is quite constant.
This tendency can be even better seen in \autoref{fig:hum_seat} (b). During the weekdays (Mo - Fr) there is a prominent increase in the occupation around lunchtime for both locations.
For \emph{V} this peak is very narrow and the occupancy is reduced to a lower level. From this, it can be concluded that square \emph{V} is mainly used for practical reasons, such as eating lunch or waiting for the train.
For \emph{M} the spike around lunchtime is also noticeable. A second utilization peak follows right after noon, indicating that this square is also frequently visited after work.
On the weekends, the lunch spike is missing at both places. At \emph{V} the utilization is reduced when compared to weekdays and for \emph{M} higher activity in the evening (after 20:00) has been recorded. 
From these results, it can be concluded that square \emph{M} is more actively used, which might be explained by its central location and attractive surroundings. Square \emph{V} on the contrary is clearly used as a stop-by location for having lunch, changing trains, or having a small break.

\begin{figure}[]
    \includegraphics[width=\columnwidth]{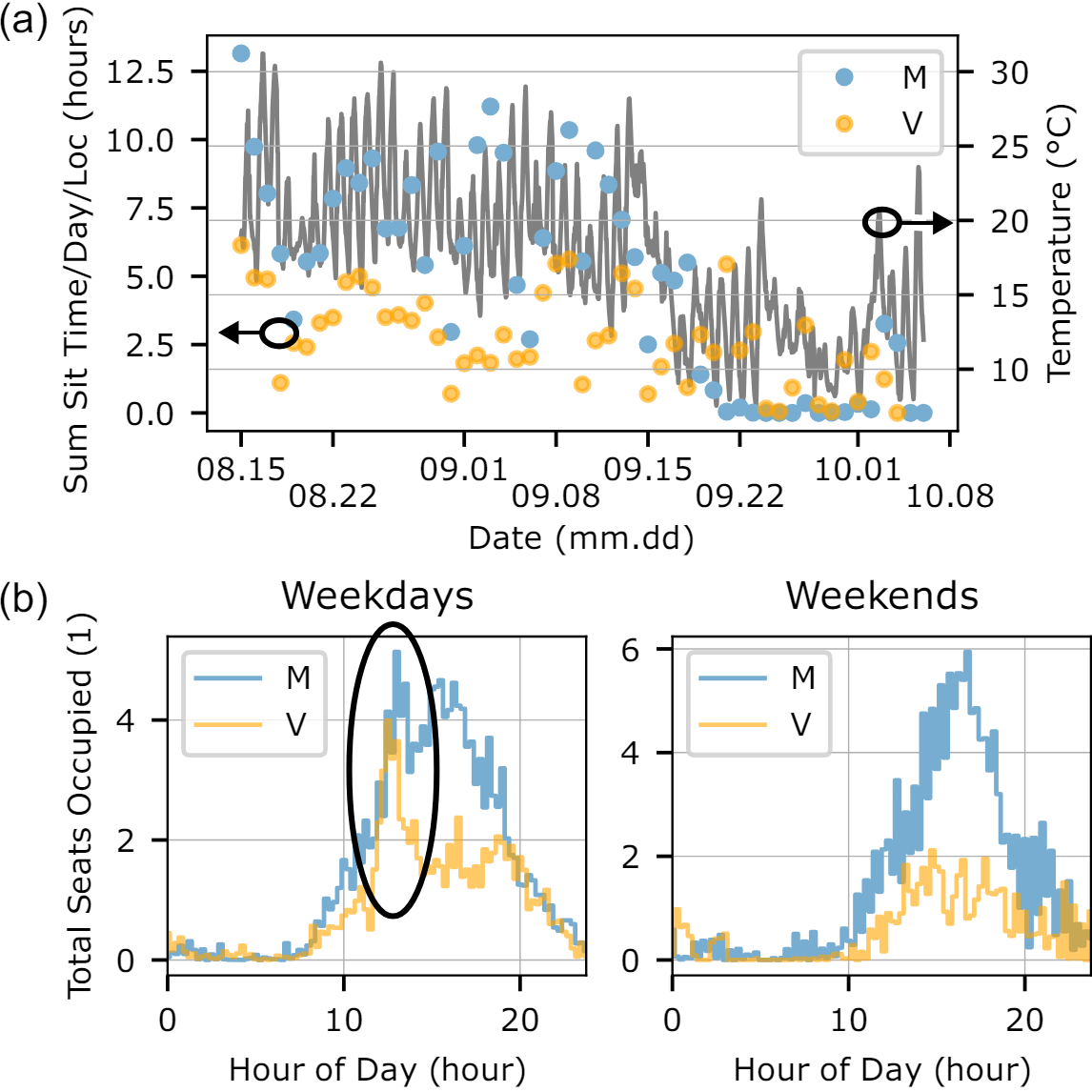}
    \caption{Recorded seat usage (Total time per day per location) for both locations compared with recorded temperature (a) and accumulated occupied seats (granularity 0.25 hours) over the 2 months with respect to hour of the day. Measurements were filtered for days of the week, accumulated to Weekdays (Mo - Fr) divided by 5 for normalization, and Weekends (Sa, Su) divided by 2 for normalization. Lunch hour is marked on Weekdays (b).}
    \label{fig:hum_seat}
\end{figure}


%% file: src/05_Conclusion.tex
\section{Conclusion}\label{sec:conclusion}
This paper presented the design, implementation, and in-field evaluation of energy-efficient long-range wireless smart sensors for urban data collection on public squares in cities to facilitate future urban data collection. 
The achieved battery lifetime from non-rechargeable batteries is 2 months, reducing maintenance and allowing simple deployment. 
In-field experimental evaluation of over 2 months of data has already shown an indication of the potential knowledge gained by employing such sensors. 
All data collected during the field trial is publicly available as open data \cite{open_data}. 


%% file: inc/acronyms.tex
\begin{acronym}
    \acro{RF}{Radio Frequency}
    \acro{IoT}{Internet of Things}
    \acro{GNSS}{Global Navigation Satellite System}
    \acro{IMU}{Inertial Measurement Unit}
    \acro{WWAN}{Wireless Wide Area Network}
    \acro{LoRaWAN}{Long Range Wide Area Network}
    \acro{LoRa}{Long Range}
    \acro{MCU}{Microcontroller}
    \acro{SoC}{System on Chip}
    \acro{I2C}{Inter-Integrated Circuit}
    \acro{PCB}{Printed Circuit Board}
    \acro{RTOS}{Real-Time Operating System}

\end{acronym}